\documentclass[fleqn,usenatbib]{mnras}

\usepackage{graphicx}
\usepackage{graphics}
\usepackage{amsmath, amssymb}
\usepackage{multirow}
\usepackage{color}
\usepackage{xcolor}
\usepackage{bm}		
\usepackage[normalem]{ulem}

\usepackage{float}

\def\kms{{\rm\,km\,s^{-1}}}
\def\kmskpc{{\rm\,km\, \,s^{-1} \, {kpc}^{-1}}}

\def\deg{{^\circ}}

\def\kpc{{\rm kpc}}

\def\mathnew{\mathsurround=0pt}   
\def\simov#1#2{\lower .5pt\vbox{\baselineskip0pt  
    \lineskip-.5pt\ialign{$\mathnew#1\hfil##\hfil$\crcr#2\crcr\sim\crcr}}}

\def\'#1{\ifx#1i{\accent"13\i}\else{\accent"13#1}\fi}    
  
\def\et{et~al. }

\def\aap{{A\&A}}

\def\araa{{ARA\&A}}
\def\aj{{AJ}}
\def\apj{{ApJ}}
\def\apjl{{ApJL}}
\def\nat{{Nature}}
\def\mnras{{MNRAS}}
\def\apjs{{ApJS}}

\def\rmxaa{{Rev. Mex. Astron. Astrofis.}}

\newcommand{\TA}{$\rm ^a$}

\title[The imprint of Arms and Bars on Rotation Curves]{The imprint of arms and bars on rotation curves: in-plane and off-plane}

\author[Martinez-Medina et al.]{Luis A. Martinez-Medina\thanks{Contact e-mail:\href{mailto:lamartinez@astro.unam.mx}{lamartinez@astro.unam.mx}}, Barbara Pichardo\thanks{In memoriam.}, 
Antonio Peimbert\\
Instituto de Astronom\'ia, Universidad Nacional Aut\'onoma de M\'exico, A.P. 70--264, 04510, M\'exico, CDMX, M\'exico
}
\date{Released \today}

\pubyear{2020}
\begin{document}
\label{firstpage}
\pagerange{\pageref{firstpage}--\pageref{lastpage}}
\maketitle

\begin{abstract}

Within Rotation Curves (RC) is encoded the kinematical state of the stellar disc as well as information about the dynamical mechanisms driving the secular evolution of galaxies. To explain the characteristic features of RCs that arise by the influence of spiral patterns and bar, we study the kinematics of the stellar disc in a set of spiral galaxy models specifically tailored for this purpose. We find that, for our models, the induced non-circular motions are more prominent for spirals with larger pitch angle, the ones typical in late type galaxies. Moreover, inside corotation, stars rotate slower along the spiral arms than along the inter-arm, which translates into a local minima or maxima in the RC, respectively. We also see, from off-plane RC, that the rotation is faster for stars that at observed closer to the plane, and diminishes as one looks farther off plane; this trend is more noticeable in our Sa galaxy model than our Sc galaxy model. Additionally, in a previous work we found that the diagonal ridges in the $V_{\phi}-R$ plane, revealed through the Gaia DR2, have a resonant origin due to the spiral arms and bar and that these ridges project themselves as wiggles in the RC; here, we further notice that the development of these ridges, and the development of high orbital eccentricities in the stellar disc are the same. Hence, we conclude that, the following explanations of bumps and wiggles in RCs are equivalent: they are manifestations of diagonal ridges in the $V_{\phi}-R$ plane, or of the rearrangement of the orbital eccentricities in the stellar disc.

\end{abstract}                

 
\section{Introduction} \label{sec:intro}
Since the early discovery of rotation of galaxies by
\citet{1914LowOB...2...66S} and \citet{1918PNAS....4...21P}, that gave
origin to the first studies on full disc rotation curves
\citep[e.g.,][]{1970ApJ...159..379R,1973A&A....26..483R,1978ApJ...225L.107R,1981AJ.....86.1791B,1981AJ.....86.1825B,1988AJ.....96..851G,1985ApJ...295..305V, 1989A&A...223...47B,1985A&A...146..213R}, they have been key tools to understanding galaxies. Rotation curves are now employed to derive the mass distribution in disc galaxies, to study discs kinematics, to study galaxy evolution (by comparing rotation curves in distant galaxies with the ones in nearby galaxies), as well as to study departures from circular rotation (i.e. the effect of non-axisymmetric large scale structure in disc galaxies, such as spiral arms and bars), among other things; for a complete review on the type of studies that can be made on rotation curves see, for example, \citet[][]{2001ARA&A..39..137S}. 

Determinations of rotation curves in disc galaxies have, in the last few decades, been seriously improved with better telescopes, instrumentation, and expertise. The main observational data used to determine rotation curves in gas and stars include: H$\alpha$ and other optical measurements, \ion{H}{1} line, CO line, maser lines, as well as planetary nebulae \citep[][and references therein]{2001ARA&A..39..137S}. Rotation curves vary depending upon the type of galaxy and, within a given galaxy, upon the tracer.

Rotation curves are very similar in shape for different galaxy types, however the maximum amplitude varies considerably for Sa, Sb, and Sc galaxies \citep[with $V_{max}$ values of approximately 300, 220, and 175 km s$^{-1}$, respectively;][]{1985ApJ...289...81R}. Albeit the gravitational potential  in the disc and halo is weakly dependent on the optical luminosity distribution  shape, there is a better correlation between total luminosity and $V_{max}$. In terms of luminosity (or mass): brighter (more massive) galaxies show slightly declining rotation curves, while less luminous (less massive) galaxies present monotonically increasing rotation curves \citep{1996MNRAS.281...27P}. Also, differences are noticeable in rotation curves at the inner regions of different galaxy types: from Sa to Sc galaxies, increasingly steeper rises and higher central velocities within a few hundred pc of the nucleus are observed \citep{1999ApJ...523..136S}; in particular, dwarf galaxies show in general smoother central rises.

Regarding the tracer, rotation curves are determined typically with atomic, ionized, or molecular gas, as well as with stars. Every tracer has its interpretation difficulties \citep{2018MNRAS.477..254L}, for example, the gas component velocity fields are complex due to their intrinsic turbulent nature, and, in the particular case of molecular gas traced through the emission of the CO molecule, is supposed to be the coldest (with dispersions of about 10 km s$^{-1}$) and, as mentioned before, is therefore supposed to trace better the circular velocity of disc galaxies; however, for this component of the interestellar medium, spiral arms and bar are influential in re-shaping the orbits \citep[e.g.][]{1999MNRAS.302L..33L,2007ApJ...665.1138S}. Stellar kinematics on the other hand, has its problems too due to their high velocity dispersion, that makes measurements noisy and complicated to interpret; circular velocities from galaxies derived from stellar kinematics can provide (rather rough) estimations of the total dynamical mass; of course, this is better calculated in function of how well the stellar velocity dispersion is considered. 

\citet{2018MNRAS.477..254L} have tested three widely used dynamical models: the asymmetric drift correction method, the Jeans method, and Schwarzschild models method; all employed to derive masses from the assumption of circular velocities from stellar kinematics for 54 galaxies observed with the CALIFA and EDGE surveys (stellar data from CALIFA and CO data from EDGE); their results show that the three models reproduce the CO circular velocity at $1R_e$(effective radius) to within 10$\%$; the three models show a large scatter (up to 20$\%$) in the inner regions, where they warn some assumptions may break down. In the same direction, \citet{2015A&A...578A..14C}, caution about the tangent-point method (extensively employed to infer velocities for the inner low-latitude regions of the Galactic disc) where they show, with a numerical simulation of the MW, that the resulting velocity profile strongly deviates from the rotation curve of the simulation due to overestimations, particularly towards the central regions, for the presence of the bar, but also in the external regions for the presence of the spiral arms. Finally, from the observational point of view, some methods have been developed to try to remove the effect of non-axisymmetric features in galaxies to obtain smooth rotation curves \citep[e.g.][]{1987PhDT.......199B,2004ApJ...605..183W}; however, they only provide rough approximations to the real mass, specially in the inner regions of barred discs.

To a first order, a typical procedure is to ignore the effects of spiral arms, and, in many cases, even the bar to study different aspects of the behavior of stars and gas in galaxies \citep[for a revision see][]{2014MNRAS.440.2564F}. In the particular case of the rotation curves this is a strong assumption on which rely the vast majority of studies, for example to determine masses of a given galaxy. Some studies however, have started to consider the spiral arms as influential features in the Milky Way Galaxy in diverse forms \citep[e.g.][]{2014MNRAS.440.2564F,2012A&A...548A.126M,2012MNRAS.426.2089R,2009ApJ...700L..78A,2011MNRAS.417..698L,2011MNRAS.417..762Q}, and also, specifically in rotation curves in galaxies in general \citep[e.g.][]{1999A&A...342..627B}. See \citet{2014RvMP...86....1S} for a review.

Serious improvements have been performed scientifically and technologically to telescopes, instruments, and computers that allow us to resolve and study with unprecedented detail the effects of spiral arms and bars. Rotation curves at different vertical heights from the Galactic plane have been derived by different authors for the Milky Way Galaxy \citep[][]{Lopez_Corredoira14,2013MNRAS.436..101W,2010ApJ...716....1B,2002ApJ...574L..39G}; in those papers authors find a nearly flat rotation curve with a small decrement for higher values of $|z|$. Finally, in a rather subtle manner, superposed on the rotation curve are smooth fluctuations, of up to tens of km s$^{-1}$, due to spiral arms. For the inner regions of barred galaxies, the fluctuations are larger, of the order of 50 km s$^{-1}$, arising from non-circular motions in the oval potential \citep{2001ARA&A..39..137S}.

In this work we explore some morphological characteristics of in-plane and off-plane rotation curves for spiral galaxies in relation to galaxy properties, such as the Hubble type, the detailed structure of the spiral arms, as well as the characteristics of the bar in an specific Milky Way application. To this purpose we employ the detailed semi-analytic galactic potential PERLAS \citep{PMME03,Pichardo2004,2012AJ....143...73P}. With this potential we adjust a set of galactic morphological types and compute millions of orbits to study in detail and extensively the behavior of stars and orbits influenced by the spiral arms (as well as the bar, in a Milky Way application); we also have produced a study of the rotation curve for stars with different ages and metallicities. This work offers important differences with respect to others where galaxies are assumed to be axisymmetric, or where the spiral arms and bar modeling is bidimensional or results simplistic.

This paper is organized as follows: in Section \ref{model} the models and simulations are described; in Section \ref{rotcur} we study the rotation curves of our spiral galaxy models; in Section \ref{rotcurZ} we study the rotation away from the mid-plane; Section \ref{MilkyWay} contains the particular case of the MW. Finally, in Sections \ref{conclusions} we present our main conclusions.

\section{Galactic mass model and Simulations}\label{model}
A key ingredient to produce a comprehensive orbital study to represent a given system, such as disc galaxies (our labor here), is the specific potential employed, in particular the non-axisymmetric components. For this work we use an elaborate, three dimensional model, adjusted to better represent structural and dynamical parameters of a given type of galaxy. Our model is steady, and we chose this over more sophisticated N-body simulations to use it in statistical orbital studies. Furthermore, the model is fully adjustable and computationally fast, and, because of its nature, this type of models allows us to study in great detail individual stellar orbital behaviour, without the known resolution problems of $N$-body simulations.

The model is composed by a three dimensional potential with disc, halo, and tumbling bisymmetric spiral arms and/or bar components (depending on the experiment). We present a brief summary of the model in this section; for a full description of the potential, see \citet{PMME03,Pichardo2004}; the updated parameters for the Milky Way case, can be found in  \citet[]{2012AJ....143...73P,2015ApJ...802..109M,2015MNRAS.451..705M,2017MNRAS.468.3615M,2018MNRAS.474...32M}, and the parameters for galaxies in general can be found \citet[][]{2015MNRAS.451.2922Pa,2015ApJ...809..170Pb,2015ApJ...802..109M,2012ApJ...745L..14P,2013ApJ...772...91P}. 

The model is composed by an axisymmetric background potential and a non-axisymmetric part. The axisymmetric potential consists of a Miyamoto-Nagai, MN, disc \citep{1975PASJ...27..533M}, a MN spherical bulge, and dark matter spherical halo \citep{AS91}. The non-axisymmetric part can include either three dimensional spiral arms, a bar, or both. 

The spiral arms potential is formed by a bisymmetric three-dimensional density distribution built of individual inhomogeneous oblate spheroids \citep[PERLAS model][]{PMME03}, placed in a logarithmic spiral locus. In this manner, the density of the arms falls exponentially along the arm axis, and for the mass distribution on the spheroids, falls linear transversal to the arms. The total mass of the spiral arms, the pitch angles, the angular velocity, and the scale-lengths used in these experiments, are presented in Table \ref{tab:nas-parameters}.

\begin{table}
\begin{center}
\caption{Parameters of the axisymmetric components of disc galaxies 
\label{tab:as-parameters}}
\begin {tabular}{lcccc}

\hline
{Parameter}&\null\hspace{5pt}Sa\hspace{5pt}\null&\null\hspace{5pt}Sb\hspace{5pt}\null&\null\hspace{5pt}Sc\hspace{5pt}\null&{Notes}\\
\hline
{M$_{B}$ / M$_{D}$} & 0.9& 0.4& 0.2 & 1,2 \\
{M$_{D}$ / M$_{H}$\TA}&0.07 &0.09 & 0.1  &  2,3  \\
{Rot. Curve ($\kms$)}& 320&250 &  170& 4  \\
{M$_{D}$ ($10^{10}$M$_\odot$)}   &12.8 & 10.0& 5.1& 3  \\
{M$_{B}$ ($10^{10}$M$_\odot$)}  & 11.6& 4.45& 1.02& $M_{B}/M_{D}$  \\
{M$_{H}$ ($10^{12}$M$_\odot$)}     &1.64 &1.25 & 0.48 & $M_{D}/M_{H}$  \\
{Disc scale-length ($\kpc$)} & 7 & 5 & 4 & 1,3\\
\hline
\multicolumn {4}{c}{\it Adopted Scales}  \\
\hline \vspace{-6pt}\\
Bulge:\hspace{10pt}   b$_1$ ($\kpc$)& 2.5 & 1.7 &1.0& \\
Disc: \hspace{12pt} a$_2$ ($\kpc$) &7.0&5.0 &4.0&\\
Disc: \hspace{12pt} b$_2$ ($\kpc$)&1.5&1.0&0.25&  \\
Halo: \hspace{12pt} a$_3$ ($\kpc$) &18.0&16.0&12.0&\\
\hline
\end{tabular}
\end{center}

\TA { Up to 100 kpc halo radius.} \\
{ References: 1)~Weinzirl \et 2009. 
                 2)~Block \et 2002.
                 3)~Pizagno \et 2005. 
                 4) ~Brosche 1971; Ma \et 2000; Sofue \& Rubin 2001.
                 
}
\end{table}

\begin{table}
\begin{center}
\caption{Parameters of the spiral arms 
\label{tab:nas-parameters}}
\begin {tabular}{lcccc}
\hline
Parameter&\null\hspace{5pt}Sa\hspace{5pt}\null&\null\hspace{5pt}Sb\hspace{5pt}\null&\null\hspace{5pt}Sc\hspace{5pt}\null&{Notes}\\
\hline
Locus                     & \multicolumn{3}{c}{Logarithmic } & 1,2,3\\
Number of arms            &  2    &  2    &  2    & 4\\
Pitch angle ($\deg$)      & 10    & 15    & 30    & 5,6 \\
M$_{sp}$/M$_{D}$          &  0.07 &  0.04 &  0.02 & 7  \\
Scale-length ($\kpc$)     &  8    &  6    &  4    & disc based\\
$\Omega_{sp}$ ($\kmskpc$) &-35    &-25    &-20    & 1,8\\
\null \hspace{20pt}(clockwise) &&&\\
ILR position ($\kpc$)     &  3.0  &  2.29 &  2.03 & \\
CR position ($\kpc$)      & 10.6  & 11.14 &  8.63 & \\
Inner limit   ($\kpc$)    &  3.0  &  2.29 &  2.03 & $\sim$ILR\\
Outer limit ($\kpc$)      & 10.6  & 11.14 &  8.63 & $\sim$CR\\ 
\hline
\end{tabular}
\end{center}
               { References:
                 1)~Grosb\o l \& Patsis 1998.
                 2) Pichardo et \et 2003
                 3)~Seigar \& James 1998; Seigar \et 2006.
                 4)~Drimmel \et 2000; Grosb\o l \et 2002; Elmegreen \& Elmegreen 2014.
                 5)~Brosche 1971; Ma \et 2000; Sofue \& Rubin 2001.
                 6)~Kennicutt 1981. 
                 7)~P\'erez-Villegas \et 2015b.
                 8)~Patsis \et 1991; Grosb\o l \& Dottori 2009; 
                      Egusa \et 2009; Fathi \et 2009; Gerhard 2011.   
}
\end{table}

The bar we selected, for the particular application to the MW case, is a triaxial inhomogeneous ellipsoid built in a similar way to the spiral arms: as a superposition of a large number of homogeneous ellipsoids to achieve a smooth density fall \citep{Pichardo2004}. The model approximates the density fall fitted by \citet{1998ApJ...492..495F} from the COBE/DIRBE observations of the Galactic centre. The total mass of the bar is $1.4 \times 10^{10}$ M$_{\sun}$, within the observational limits. A long list of studies have estimated the angular velocity of the bar, concluding that the most likely value lies in the range $\Omega = 45$ - $60$ km s$^{-1}$ kpc$^{-1}$. We adopt the lower value, $\Omega = 45$ km s$^{-1}$ kpc$^{-1}$, based on the formation of moving groups in the solar neighbourhood \citep{2009ApJ...700L..78A}. For a detailed list of the parameters adopted here to simulate the case of the MW (including both the axisymmetric and non-axisymmetric components), see Section 2 of \citet{2017MNRAS.468.3615M}. 

\section{Rotation Curves in Spiral galaxies}\label{rotcur}

The presence of non-axisymmetric structures, like spiral arms and bar, implies that within the disc, the underlying density and gravitational potential depend on the angular position, which in turn, may be reflected on the kinematics, and hence, the rotation curve at that position.

\begin{figure*}
\begin{center}
\includegraphics[width=18cm]{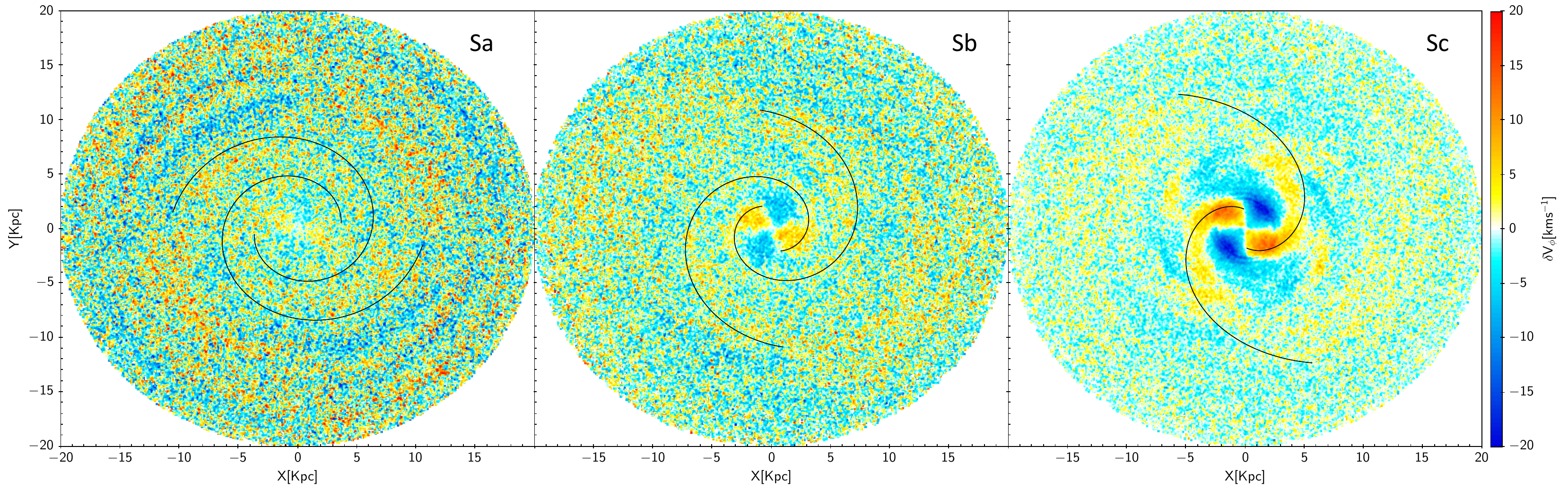}
\end{center}
\caption{Face-on map of the residual rotation velocity, defined as $\delta V_{\phi}(R,\phi) = V_{\phi}(R,\phi) - \bar{V}_{\phi}(R)$, for the three galaxy models, where $\bar{V}_{\phi}(R)$ is the azimuthal average of $V_{\phi}(R,\phi)$. Notice that galaxies rotate clockwise, while the angle increases anti-clockwise, therefore red means slower than $\bar{V}_{\phi}(R)$ and blue means faster than $\bar{V}_{\phi}(R)$ (see text).}
\label{residuals}
\end{figure*}

\begin{figure*}
\begin{center}
\includegraphics[width=18cm]{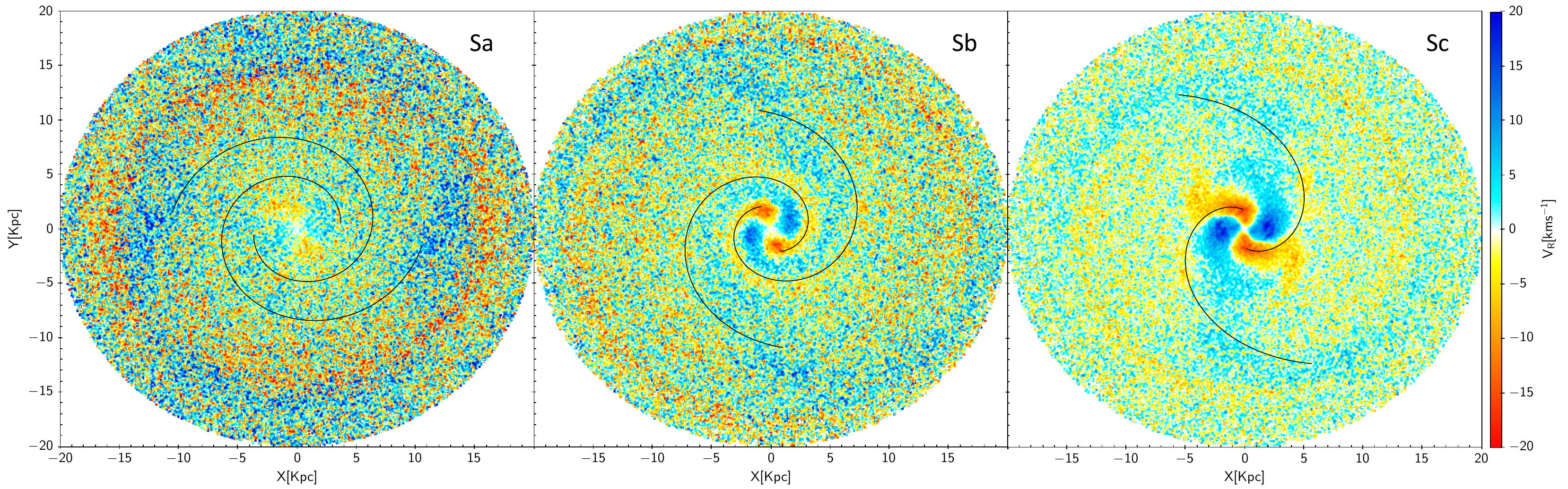}
\end{center}
\caption{Face-on map of the radial velocity for the three galaxy models.}
\label{radial}
\end{figure*}

A straightforward measurement to reveal the influence of spiral arms in the kinematics of the stellar disc is to compute the velocity maps for each of the galactic models.  Figures \ref{residuals} and \ref{radial} show the residual rotational velocity and radial velocity, respectively for our Sa, Sb, and Sc models. Non-circular motions are clearly induced in the stellar disc, which correspond to the underlying spiral overdensity and even propagate to the outer disc. 

Figure \ref{residuals} shows that, at each radius, stars along (or slightly behind) the spiral arm rotate slower than the stars in the inter-arm (or slightly ahead of the arm); this behavior can be seen more clearly in the Sb and Sc models, where the larger pitch angles produce more violent encounters with the stars.Since we are inside corotation, this can be interpreted as stars being stalled for a bit as they cross the spiral overdensity. This means that, spiral arms with steeper pitch angles would present less luminous inter-arm regions and brighter spiral arms, i.e., larger luminosity (or density) contrasts between the arm and the inter-arm regions are produced by the stalling of the stars in the spiral arms. In section 3.1 we will describe an additional kinematic mechanism to enhance the brightness along the spiral arms.

Meanwhile Figure \ref{radial} shows that, because of the presence of the spiral arms, stars on the spirals have radial velocities pointing toward the galactic center, while those in the inter-arm region have a radial velocity pointing away from the center. This, together with the results from Figure \ref{residuals}, show the kind of circulation expected from epicycles. This phenomenology is clearer in late type galaxies than in early types since smaller pitch angles produce smaller perturbations, allowing nearly circular orbits.

Once we know that the spiral arms alter in a clear way the kinematics of the stellar disc, we analyse the imprint of those non-circular motions on a global observable as is the rotation curve, which we measure along different azimuths over the disc plane.

\begin{figure}
\begin{center}
\includegraphics[width=8cm]{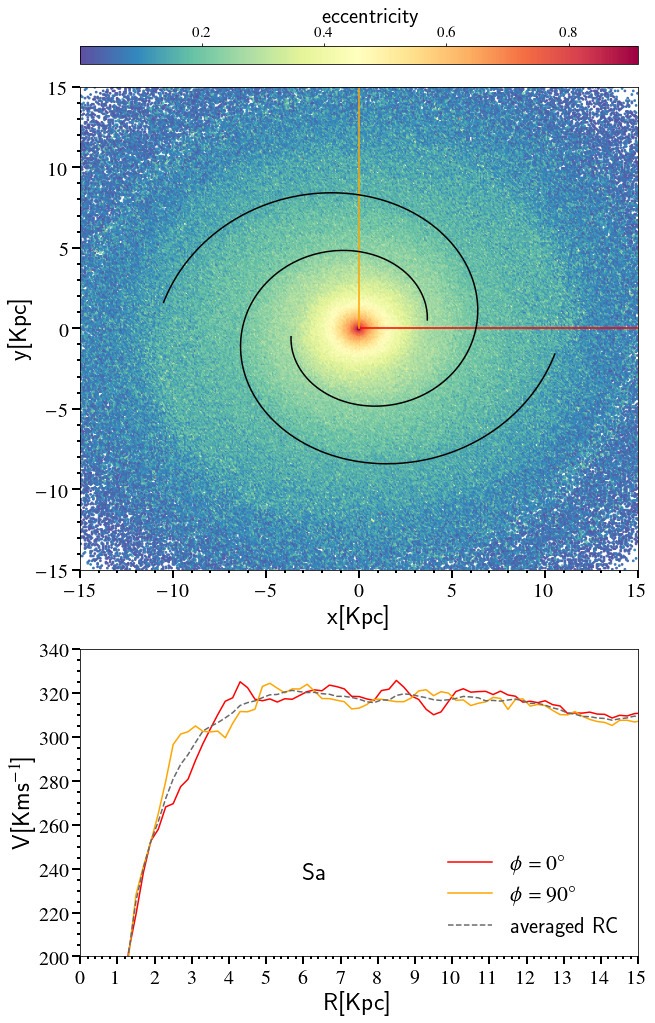}
\end{center}
\caption{Top: Eccentricity map of the Sa galaxy disc. The color coding indicates the eccentricity for each orbit, and white lines indicate the orientation of the spirals. Bottom: Rotation curve measured along $\theta = 0{\deg}$ (red curve) and $\theta = 90{\deg}$ (orange curve) as indicated in the top panel; also included is the azimuthal averaged rotation curve (dashed line).}
\label{Sa}
\end{figure}

For this analysis we take each of the simulations at $t = 2Gyr$. Figure \ref{Sa} (top panel) shows the $x - y$ projection of the Sa stellar disc and the locii of the spiral arms. The color assigned to each particle corresponds to its orbital eccentricity $e$, which is computed by tracking back each orbit during the 500 Myr prior to their current position; (We choose this because it leaves most of the time of the simulation for the orbits to approach equilibrium, yet it allows us to trace them for enough time so most particles can go around the galaxy at least once.).  This allows to record the value of the apocenter $R_a$ and pericenter $R_p$ for each orbit, and we express their eccentricity as:
\begin{equation}
\label{ecc}
    e = \frac{R_a-R_p}{R_a+R_p}.
\end{equation}
Then, we draw two radial lines at $\theta = 0{\deg}$ and $\theta = 90{\deg}$ as indicated in the top panel of Figure \ref{Sa}; meanwhile the bottom panel shows the rotation curve computed for stars that lie along a $10\deg$ arc centered on each of those two lines and with $e \leq 0.3$. The curves present local minima and maxima that are clearly different for both lines, they even show significant anticorrelation. The line at $\theta = 90{\deg}$ crosses the spiral patten twice, at $\sim 4kpc$ and $\sim 8kpc$, at those positions the curve develops local minima. Meanwhile the line drawn at $\theta = 0{\deg}$ crosses the spiral patten just once, at $\sim 6kpc$, at that radius the value of the rotation velocity is also within a minima. The extra wiggles appearing on the curves are due to the substructure that the spirals induce on the eccentricity map (top panel), which in turn reflects on the rotation curves as wiggles. 

\begin{figure}
\begin{center}
\includegraphics[width=8cm]{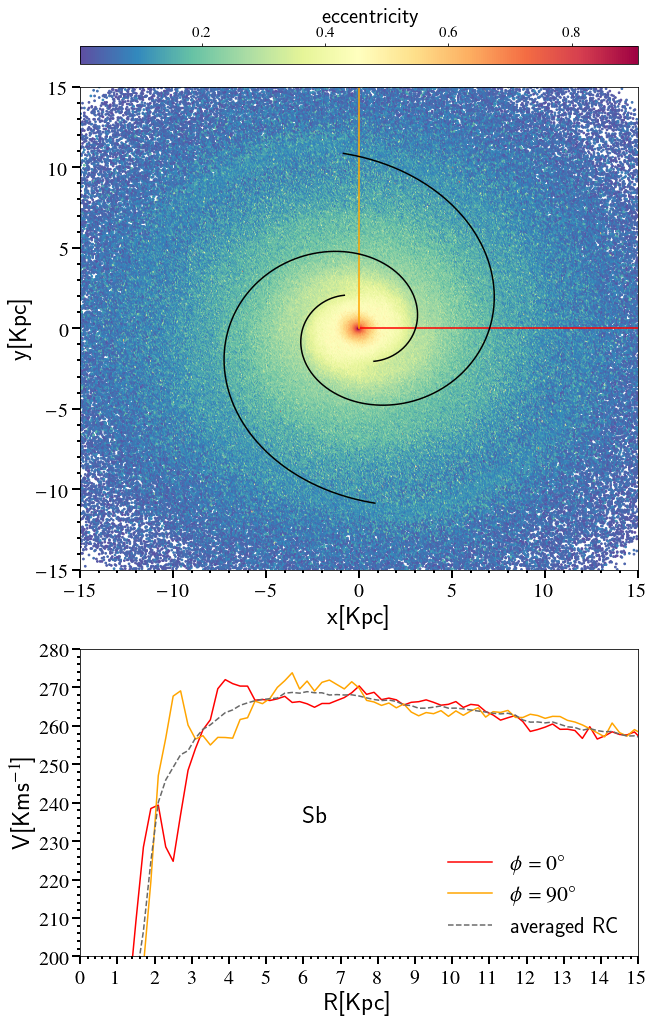}
\end{center}
\caption{Top: Eccentricity map of the Sb galaxy disc. The color coding indicates the eccentricity for each orbit, and white lines indicate the orientation of the spirals. Bottom: Rotation curve measured along $\theta = 0{\deg}$ (red curve) and $\theta = 90{\deg}$ (orange curve) as indicated in the top panel; also included is the azimuthal averaged rotation curve (dashed line).}
\label{Sb}
\end{figure}

\begin{figure}
\begin{center}
\includegraphics[width=8cm]{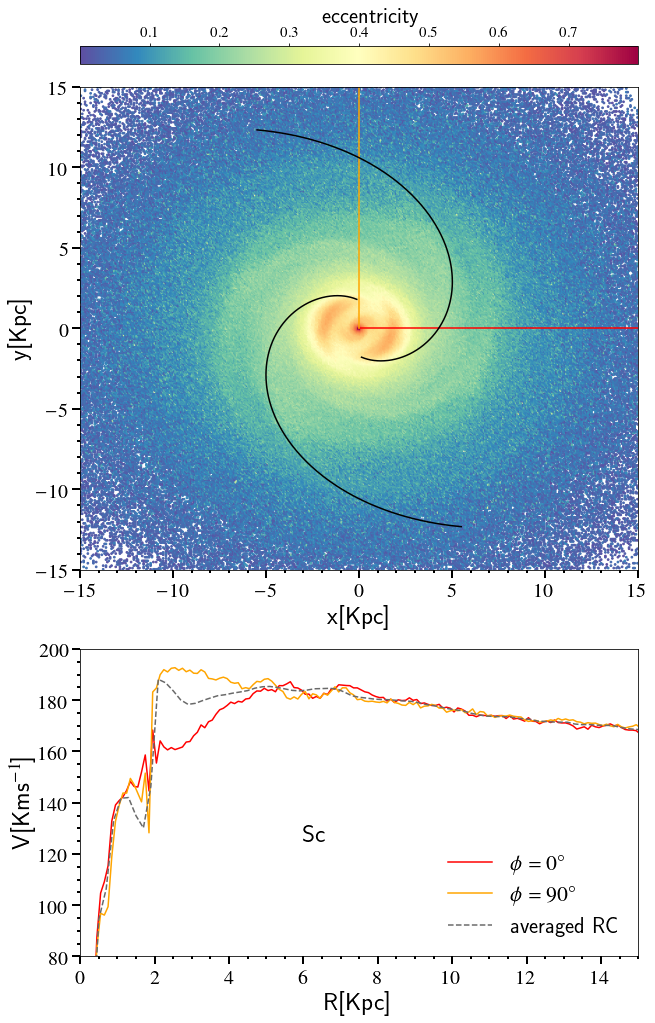}
\end{center}
\caption{Top: Eccentricity map of the Sc galaxy disc. The color coding indicates the eccentricity for each orbit, and white lines indicate the orientation of the spirals. Bottom: Rotation curve measured along $\theta = 0{\deg}$ (red curve) and $\theta = 90{\deg}$ (orange curve) as indicated in the top panel; also included is the azimuthal averaged rotation curve (dashed line).}
\label{Sc}
\end{figure}

We repeat the same analysis for the discs in the Sb and Sc galaxy models (Figures \ref{Sb} and \ref{Sc}), and the results are similar. The rotation curves measured along $\theta = 0{\deg}$ and $\theta = 90{\deg}$ are different, they develop a local minima every time the radial line crosses a spiral arm. Extra wiggles are present in the curves, that may not correspond to the position of the arms but are a direct consequence of the orbital eccentricities rearranged by the arms. 

\begin{figure}
\begin{center}
\includegraphics[width=8cm]{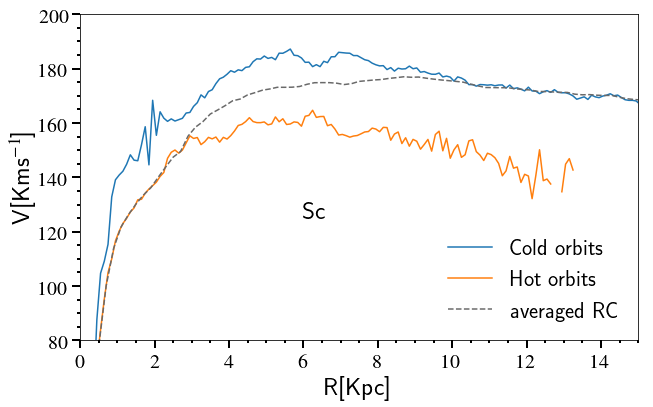}
\end{center}
\caption{Rotation curves for the cold and hot stellar populations in the Sc galaxy model, both measured along the same angle $\theta = 0{\deg}$ (blue line in top panel of Figure \ref{Sc}); also included is the azimuthal averaged rotation curve (dashed line).}
\label{ScCold}
\end{figure}

Finally, for the Sc galaxy model, we compare the rotation curves defined by highly-eccentric vs moderately-eccentric orbits; we measure the rotation curve along only one direction ($\theta = 0{\deg}$ as indicated in Figure \ref{Sc}) but for two stellar populations classified as cold $e  \leq 0.3$, and hot $e > 0.3$, respectively. Figure \ref{ScCold} shows this comparison; as expected, the rotation curve from stars in more eccentric orbits is slower, while both curves develop wiggles. Particularly the rotation curve of the cold population has a local minima at $\sim 6.4kpc$; but notice, from top panel of Figure \ref{Sc}, that at such radius the measurement line does not cross any of the arms. The reason there is a wiggle at that position is because the arms have arranged the orbital structure such that at $~6.4kpc$ there is a gap where the mean eccentricity drops before going up again. Also included is the azimuthal averaged roation curve; this curve is very similar to the hot RC at radii smaller than 3kpc, it then slowly transitions until at 10 kpc it becomes very similar to the cold RC. This transition is mostly because the fraction of hot orbits is very large in the central part of the galaxy (88\% at 3kpc) and it slowly diminishes toward the outer parts of the galaxy (5\% at 10kpc); the remaining difference comes about because the dotted line represents particles in all azimuthal directions while the colored lines represent orbits within a few degrees.

This result shows that the spiral arms left their imprints on the rotation curve, not only at their instantaneous position, but along all the area swept by them.

The eccentricity map becomes less homogeneous in the Sc galaxy model, which is the one with wide open spiral arms. We have noticed that the spiral arms rearrange orbital eccentricities in that way implies that the value of the rotation velocity depends on whether the measurement takes place in the arm or in the inter-arm, as we will show in the next section.

\subsection{Rotation along the Arm \& Inter-Arm}

In this section we measure explicitly the stellar rotation along the spiral arms and compare it with the rotation along the inter-arm region. 

\begin{figure*}
\begin{center}
\includegraphics[width=17cm]{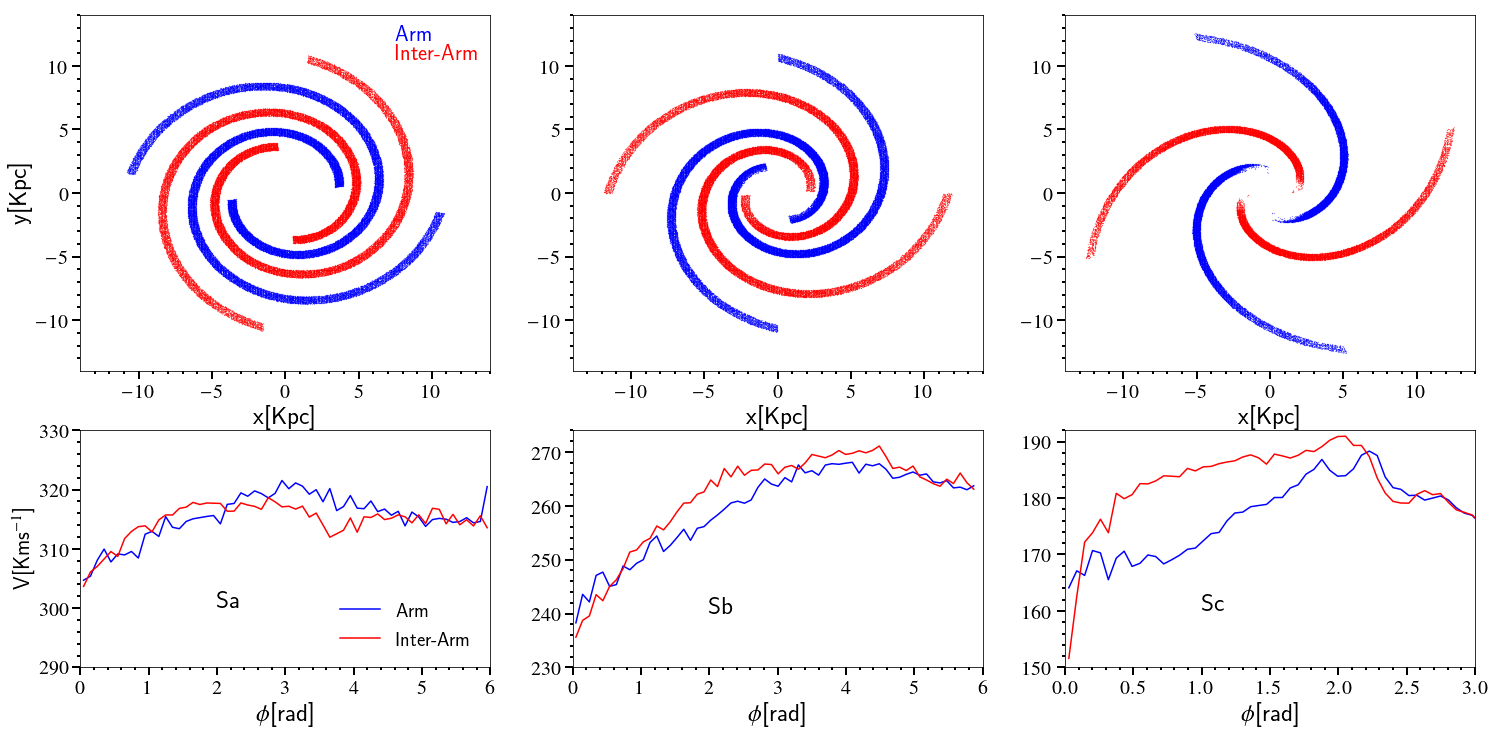}
\end{center}
\caption{Top row: For the three galaxy models, selected particles that belong to the arms (blue dots) and selected particles that belong to the inter-arm (red dots). Bottom row: Rotation curve as a function of the angle along the arms (blue line, measured on the blue dots above) and as a function of the angle along the inter-arm (red line, measured on the red dots above).}
\label{ArmInter}
\end{figure*}

The measurement of the rotation curve was done as follows: for $t = 2Gyr$ in the simulation, we compute the exact location of the spiral over-densities, and hence, also the location of the inter-arm regions. For the arms we select a spiral band of 0.6 kpc across (see the upper panels of Figure \ref{ArmInter}), narrow enough to assure it is well inside the arms, then we compute the rotation of all those particles within the selected region. After locating the stars that belong to the arms, their rotation speed is computed as a function of the angular position of the center of the spiral arm at a given radius, which starts from zero at the origin of the spiral. For the inter-arm region we follow the same procedure, only shifting the spiral bands by 90$\deg$. The procedure is better illustrated in the upper panels of Figure \ref{ArmInter}, which show the position of those particles within the arm (blue dots) and within the inter-arm (red dots) for our three galactic models.

Figure \ref{ArmInter} (bottom panels) shows a comparison of the rotation velocity along the arms and the inter-arm regions; the colors correspond to the regions plotted in the upper panels. It is important to say that blue and red bands follow a logarithmic spiral locus, covering the same radii; which means that, when comparing the two curves of each galaxy, the same angle corresponds to the same radius. First, we notice that, for the Sa galaxy, the rotation along the arms is very similar to the rotation along the inter-arm regions; however, for the Sb galaxy the two rotation curves are clearly different, although the differences are small, the rotation along the arm is systematically $3-5\kms$ smaller than the rotation along the inter-arm (2-3\% smaller); finally, for the Sc galaxy the differences between rotation curves are even larger, up to more than $15\kms$ (up to 8\%).

That the stars rotate slower along the spiral arms than along the inter-arm can be explained as stars being, in average, closer to their apogalacticon (the point of the orbit with the smallest velocity), while stars on the interarm band are, on average, closer to their  perigalacticon (the point with the largest velocity).

Finally, notice that the differences in rotation along the arm and inter-arm increase when going from Sa to Sc galaxies. Tightly wound arms do not have major influence in the orbital eccentricities, showing no significant dependence of the rotation as a function of the azimuthal angle. Meanwhile, wide open arms will induce major non-circular motions in the orbits, making their rotation depend on the azimuthal position. This, of course, depends on the mass of the spiral arms; models with different mass ratios might find a different correlation, but we consider that our choice for the masses is well founded on galactic orbital studies \citep[e.g.][]{2015ApJ...809..170Pb}.

\section{Rotation as a function of distance from the plane}\label{rotcurZ}

Up to now we have classified the stellar orbits according to their radial heating (i.e. according to their eccentricity). Since we have the 3D kinematical data of the orbits, we also can describe the stellar disc according to the vertical heating of the orbits.

As in section 3, we take a snapshot of the simulation at $t = 2$Gyr and, for every particle, we trace back its orbit for the $500$Myr, prior to the current time. During this period of time we compute the maximum vertical amplitude of the orbit $z_{max}$. In this section we refer to a stellar population as vertically cold (or hot) if the stellar orbits have small (or large) values of $z_{max}$.

\begin{figure*}
\begin{center}
\includegraphics[width=19cm]{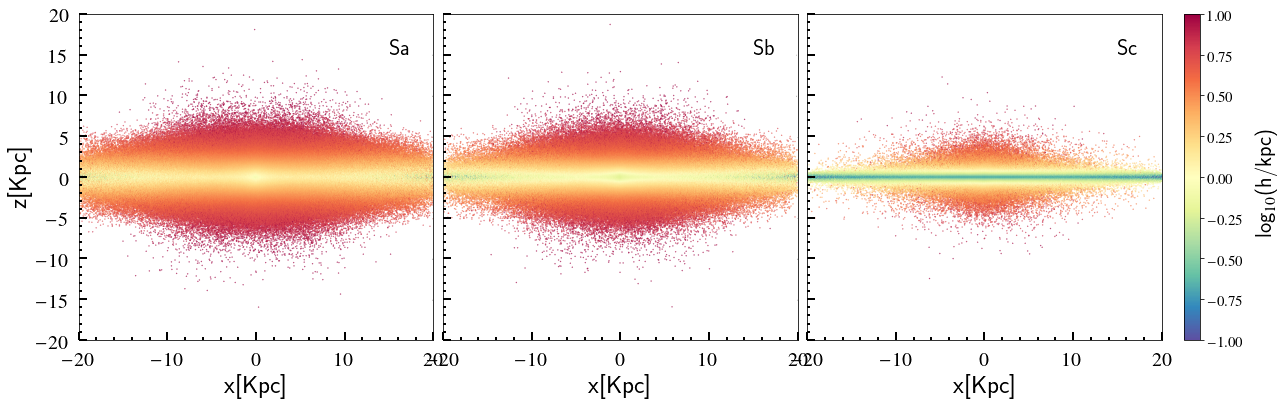}
\end{center}
\caption{Edge-on projection of the stellar disc for our three models. The color coding indicates the maximum vertical amplitude $z_{max}$ for the orbit of each particle. By classifying the orbits in this way we see that the mid-plane is dominated by a vertically cold stellar population, while far away from the plane the discs are hot. In general the entire orbital structure is hotter for a Sa galaxy and colder for the Sc galaxy.}
\label{vertmap}
\end{figure*}

Figure \ref{vertmap} shows the results of this classification for the three galactic models. It is an edge-on projection of the stellar disc, where the color coding indicates the maximum vertical amplitude $z_{max}$. It is clear that the mid-planes of the galaxies are dominated by stars with small vertical excursions, while farther away from the plane the discs are populated by vertically hot orbits. This behaviour is the same for the three galactic models, their stellar discs are composed by populations with different kinematics. However, notice that the stellar populations are hotter for the Sa galaxy, and cooler for the Sc galaxy. This is because the discs for Sa galaxies are massive, with large vertical scale lengths, hence in order to be supported, the vertical velocity dispersion should be large. On the other hand, the discs for Sc galaxies are less massive and much thinner; therefore, they can be supported by small velocity dispersion, which makes the stellar populations mostly cold.

\begin{figure}
\begin{center}
\includegraphics[width=8cm]{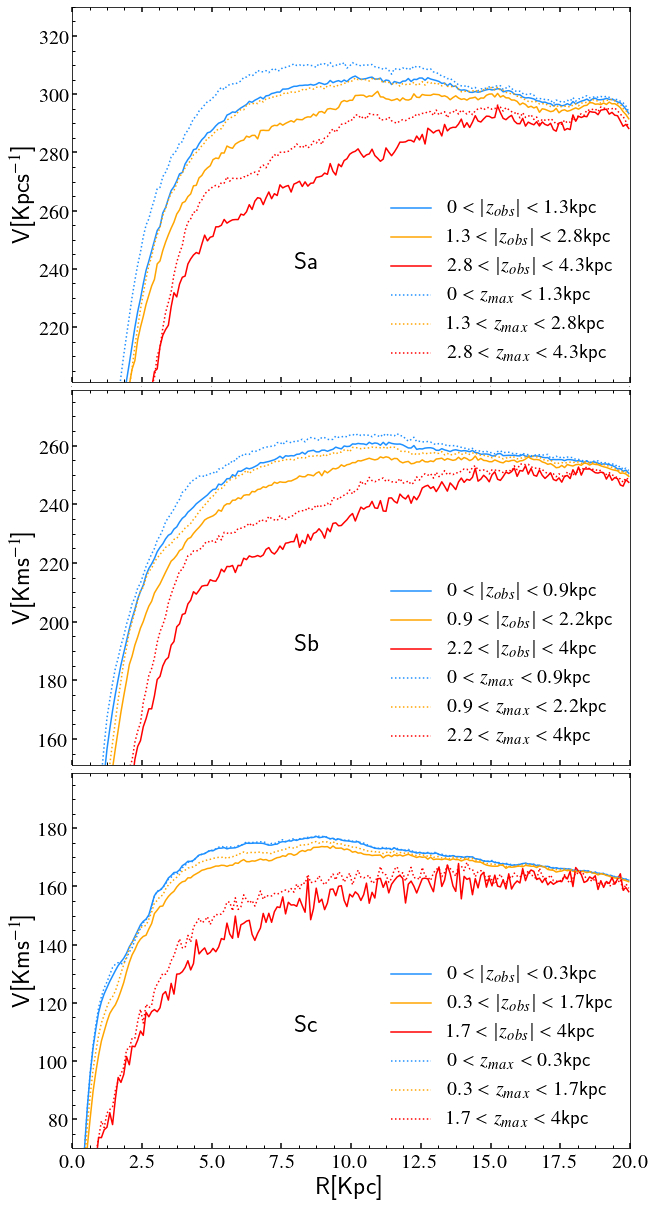}
\end{center}
\caption{Rotation curve for three stellar populations characterized by their instantaneous vertical position $z_{obs}$ (solid lines),and rotation curves of stars according to their dynamical heating $z_{max}$ (dotted lines). Note that the velocity curve defined by $|z_{obs}|$ is always above the one defined by $z_{max}$, see text.}
\label{RcVertHeat}
\end{figure}

We compute the rotation curve for stellar populations characterized only by their observed vertical height $|z_{obs}|$. In Figure \ref{RcVertHeat} we show, in solid lines, the rotation curve for three populations with different ranges of $|z_{obs}|$. To study the rotation curve at different heights above the galactic plane We define 3 vertical zones; the first zone contains the $\sim$68\% of particles above an below the mid-plane; the second zone is above/below the first zone and contains the $\sim$27\% of particles; the third zone is the one at a higher/lower altitude above the plane and contains the $\sim$4\% of particles; we are excluding the top/bottom-most 1\% of the particles. Note that the limits of the ranges are different for each of our three model galaxies. We can see that the rotation is larger for stars that at observed closer to the plane, and diminishes as the selected stars are farther away from the plane. This effect is more noticeable for Sa galaxies than for Sc Galaxies, in particular for Sc galaxies there is very little difference for the rotational velocities of stars in the 68 and 95 percentiles.

For some theoretical studies it is more important the maximum height a star can achieve rather that the height it currently possesses (e.g.  for a star to be considered part of the disc of a galaxy, it is not enough for its $|z_{obs}|$ to be small, it is also important that its $|z|$ is small at all times, i.e. its $z_{max}$ also needs to be small). For that reason we have classified the stellar populations of the disc according to their vertical heating. Figure \ref{RcVertHeat} shows, in dotted lines, the rotation velocity as a function of the galactocentric radius for three stellar populations characterized by different values of $z_{max}$. We use the same ranges for $z_{max}$ as we used for $|z_{obs}|$. Notice, again, that the coldest population, the one that dominates the plane, rotates faster. And the rotation of the orbits decreases as they reach higher vertical amplitudes. This happens for our three galactic models, and the differences in rotation between the cold and hot populations are, again, bigger in the Sa galaxy.

Overall the velocity differences for Sa galaxies are larger than for Sc galaxies (when comparing the 68 and 99 percentile), with the Sb laying somewhere in the middle; however, when considering it as a fraction of the overall velocity the Sc Galaxy has a larger  difference than both Sb and Sa galaxies (that have similar fractional differences). Another interesting trend is that Sc galaxies show very little difference for the rotational velocities of stars in the 68 and 95 percentiles, while Sa and Sb show a similar ---and larger--- deviation.

It is also interesting to compare the rotation between populations classified by their $z_{max}$ with the ones classified by their $|z_{obs}|$. We can do this by comparing the solid vs. dotted lines ($|z_{obs}|$ vs. $z_{max}$) in Figure \ref{RcVertHeat}. Notice that for all models, and for each value of the vertical separation, the population classified by $z_{max}$ rotates faster than the one classified by $|z_{obs}|$. This behaviour is due to the inherent difference in classifying according to $z_{max}$ or $|z_{obs}|$, even if both have the same value. A population classified as cold due to a small value of $|z_{obs}|$ represents the state of the mid-plane at a given time and always includes vertically hot orbits that, at that precise time, happen to be close to the plane. Meanwhile a population with a small $z_{max}$ excludes those orbits that eventually will perform large vertical excursions. This means that even if the stars lie close to the mid-plane, they may not trace properly the mean rotation of the disc; a measurement of the stellar rotation curve near to the plane always carries a contribution from vertically hot orbits (i.e. one must not forget that the rotation of the disc is not the same as the rotation on the mid-plane).
    
Notice that the differences in rotation between populations classified by the value of $z_{max}$ and by the value $|z_{obs}|$ are less important for the thinner and less massive Sc galaxy. 

Finally in the case of rotation near the plane: for the Sc galaxy model the rotation of all stars is practically the same as the rotation of the cold population; while for the Sa galaxy those differences are larger and can be as large as $5\%$.

\section{The Case of the Milky Way}
\label{MilkyWay}

Albeit we are located in a difficult position in our galaxy to dilucidate  clearly its structure and dynamics, it is still the closest and best laboratory we have to perform studies in detail; for example, its rotation curve, particularly toward the central regions, where several components take a fundamental part shaping it.

The rotation curve of the Milky Way Galaxy \citep{Clemens85,Blitz79,Brand93,Fich89,Merrifield1992,Bovylev08,Honma97,Bovy_etal2012,Reid14,Sofue09,Lopez_Corredoira14} seems to be typical of an Sb type: a non-zero velocity very close to the center (probably produced by a high-density core), a steep rise within the central 100 pc, a maximum in the rotation curve at a few hundred pc, and a minimum at 1 to 2 kpc, after which appears a gradual rise for a few kpc, and finally, a nearly flat outer rotation curve \citep{2001ARA&A..39..137S}. However the MW is a barred galaxy and the bar affects the shape of the rotation curve \citep{2016A&A...594A..86R,Valenzuela07, Spekkens_Sellwood07,Dicaire08}; for example, it seems to make it less steep toward the central parts \citep{2015A&A...578A..14C}, contrary to previous statements in which non-circular motions and specifically the orientation of the bar were not able to explain the steep nuclear rise towards the center of massive galaxies \citep{2001ARA&A..39..137S}. Barred galaxies show velocity jumps, on the leading edges of the bars, of tens and up to more than a hundred km/s; the presence of the bar makes them kinematically much more complex than non-barred galaxies. It is therefore very important, when infering any quantitative conclusion of a given galaxy from rotation curves (e.g. the mass), to consider properly non-circular motions induced by the bar \citep[e.g.][]{Spekkens_Sellwood07, Dicaire08}

In the previous section we show the behaviour of the galactic rotation curve in a stellar disc perturbed by a spiral structure. The significance of the imprint left by spiral arms on the stellar rotation depends on the galaxy morphology, and it is better understood through an analysis of the orbital eccentricities driven by the arms. We have shown this analysis for representative models of Sa, Sb, and Sc galaxies. In this section we apply the same strategy to a detailed model of the Milky Way Galaxy.

For the mass model of the Milky Way we include spiral arms and a central bar. The adopted masses and scale lengths for the different components in the model are as described in Section \ref{model}. We integrated $10^6$ orbits within our MW potential, and analysed the simulation after a $1Gyr$ evolution (this simulation has less particles and was integrated for a shorter period of time, than the simulations of the previous section, because the bar-spiral potential is much more CPU intensive than the spiral potential used in the previous section). 

\begin{figure}
\begin{center}
\includegraphics[width=8cm]{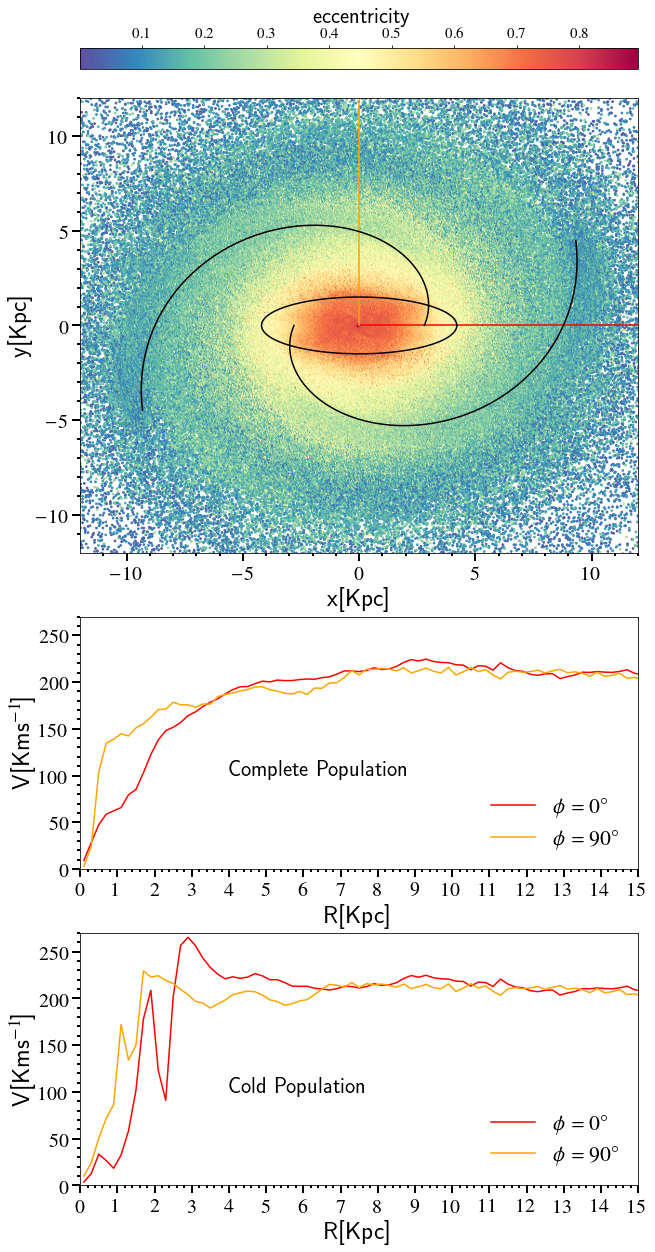}
\end{center}
\caption{Top: eccentricity map of the MW stellar disc model, white lines indicate the position of the underlying spiral and bar potentials. Middle: Rotation curve, as a function of galactocentric radius, of all the stars that lie along the lines marked in the panel above. Bottom: Rotation curve, as a function of galactocentric radius only for the stellar population with $e \leq 0.3$. }
\label{MW}
\end{figure}

Figure \ref{MW} (top panel) shows the eccentricity map at the end of the $1Gyr$ simulation; we follow each orbit for the final 300 Myr of its evolution  and use equation\ref{ecc} to obtain its eccentricity before the current snapshot. Figure \ref{MW} (top panel) also indicates the configuration of the spiral arms and bar at that moment in the simulation. Notice that the presence of both bar and spirals, result in regions of low and high eccentricities. The orbits of high eccentricities lie near the leading edge of the arms, while the arms lie along regions of relativelly low eccentricity. On the other hand, the eccentricity pattern induced by the bar is more axisymmetric, dominated by high eccentricities within the area of the disc swept by the rotating bar. The fact that the high eccentricities induced by the bar do not follow a bar-like pattern is likely due to its faster rotation compared to the rotation of the arms.

Aside from the orbital eccentricities, Figure \ref{MW} shows how bar and spirals affect the rotation curve. Middle and bottom panels show the rotation curve as a function of galactocentric radius, measured along perpendicular lines as indicated in the figure. The angular directions of these measurements were chosen such that one is parallel to the semi-major axis of the bar, while the other is parallel to the semi-minor axis.

The middle panel of Figure \ref{MW} shows the rotation velocity of all the stars that lie along each of those two chosen directions. Notice that the rotation measured along the semi-major axis of the bar is lower for all radii inside the bar. This is because the region delimited by the bar is populated mostly by $x_1$ orbits, which are aligned with the bar. This means that inside the bar the orbits have their apocenter aligned with the semi-major axis, and hence they pass through this point with their smallest velocity. The same plot also shows the imprint of the spiral arms on the rotation curve. In the direction $\phi = 90{\deg}$, the curve has a local minima between 5 and 7~kpc; it is because at those radii the measurement line passes through a region of high eccentricity, meaning a lower rotation velocity.  

Meanwhile the bottom panel of Figure \ref{MW} shows the rotation velocity only for the cold population, i.e., stars with $e \leq 0.3$. Notice that this new set of curves follow the same trends at every radius compared with the complete stellar population, but it is clear that in the cold population the features explained above are amplified. Radially cold orbits trace better the differences in the rotation curve when it is measured along different angles in a galaxy that harbours non-axisymmetric structures, like bar and spiral arms.

\subsection{Rotation as a function of age and metallicity}

\begin{figure}
\begin{center}
\includegraphics[width=\columnwidth]{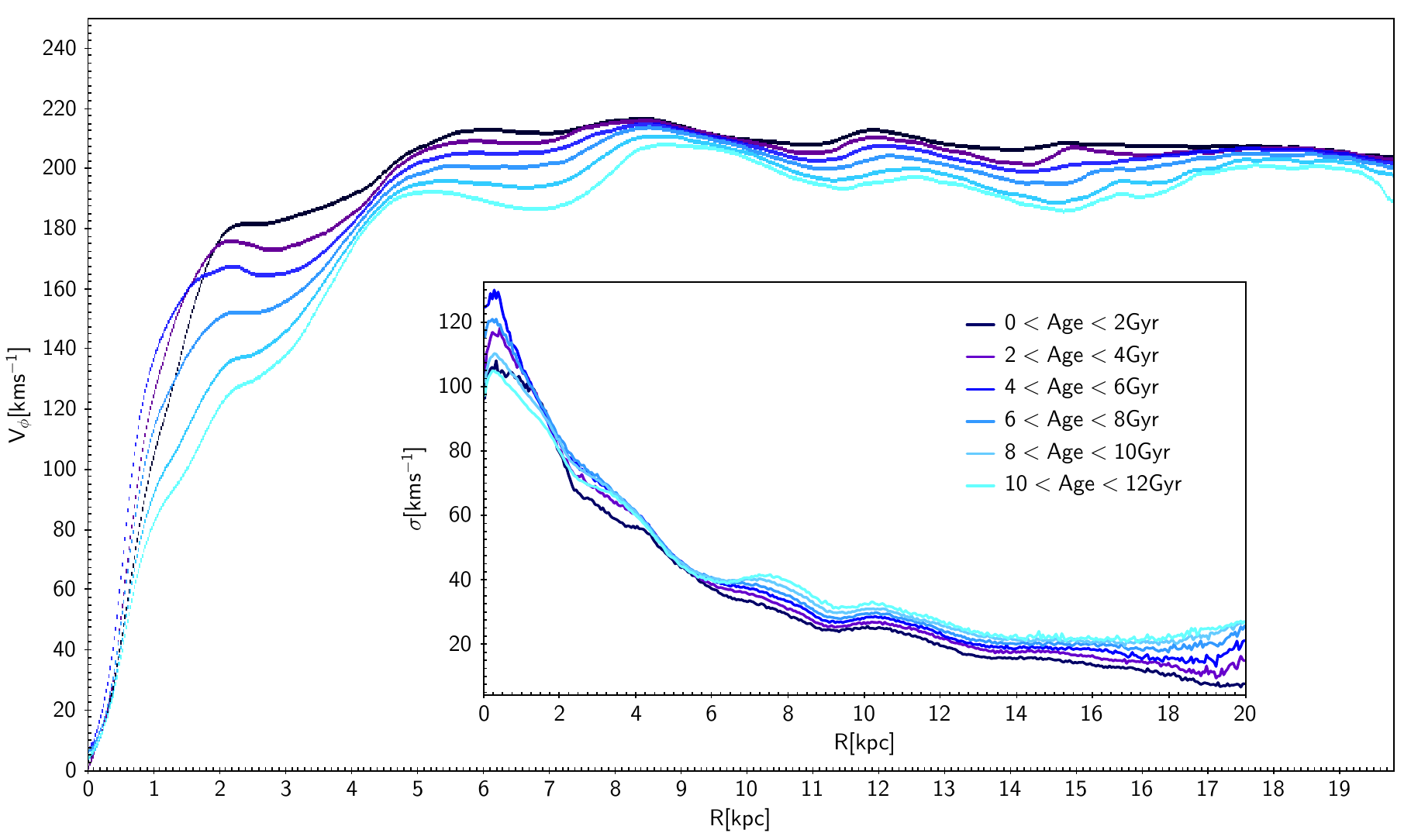}
\end{center}
\caption{Rotation velocity as a function of radius for populations with different ages.}
\label{Age2}
\end{figure}

So far we have seen that, due to the presence of spiral arms, the stellar rotation velocity depends not only on the galactocentric distance but also on the azimuthal position within the disc. But how significant the induced non-circular motions can be depends, also, on how long the star has been orbiting the disc in the presence of the spirals, i.e., on the age of the star. The simulation analysed in this section, specifically tailored for the Milky Way, assigns an age label to each star based on the time they are introduced to the simulation \citep{2017MNRAS.468.3615M}. This allows us to split the disc into populations according to their stellar age. Figure \ref{Age2} shows that the rotation does depend on the stellar age: it is larger for the younger stars and decreases with age. Similarly the velocity dispersion is larger in the old population, and decreases for the younger stars. This means that old stars are scattered  farther away from the average rotation, while the young ones have rotation velocities that are closer to the mean. Finally, notice that the amplitude of the oscillations in the RC (wiggles) is larger for the oldest population, while for the youngest one the RC develops only small perturbations.

Since the kinematics of stars is altered by non-axisymmetric structures in the disc, and the magnitude of this effect depends on the age of the star, it is expected that stars lose kinematic track of their birth place. However, a stellar property that reveals the galactocentric distance of stars at the moment of formation is their metallicity. In this way the local metallicity of the disc may change due to stars being redistributed radially and azimuthally across the disc. 

\begin{figure}
\begin{center}
\includegraphics[width=\columnwidth]{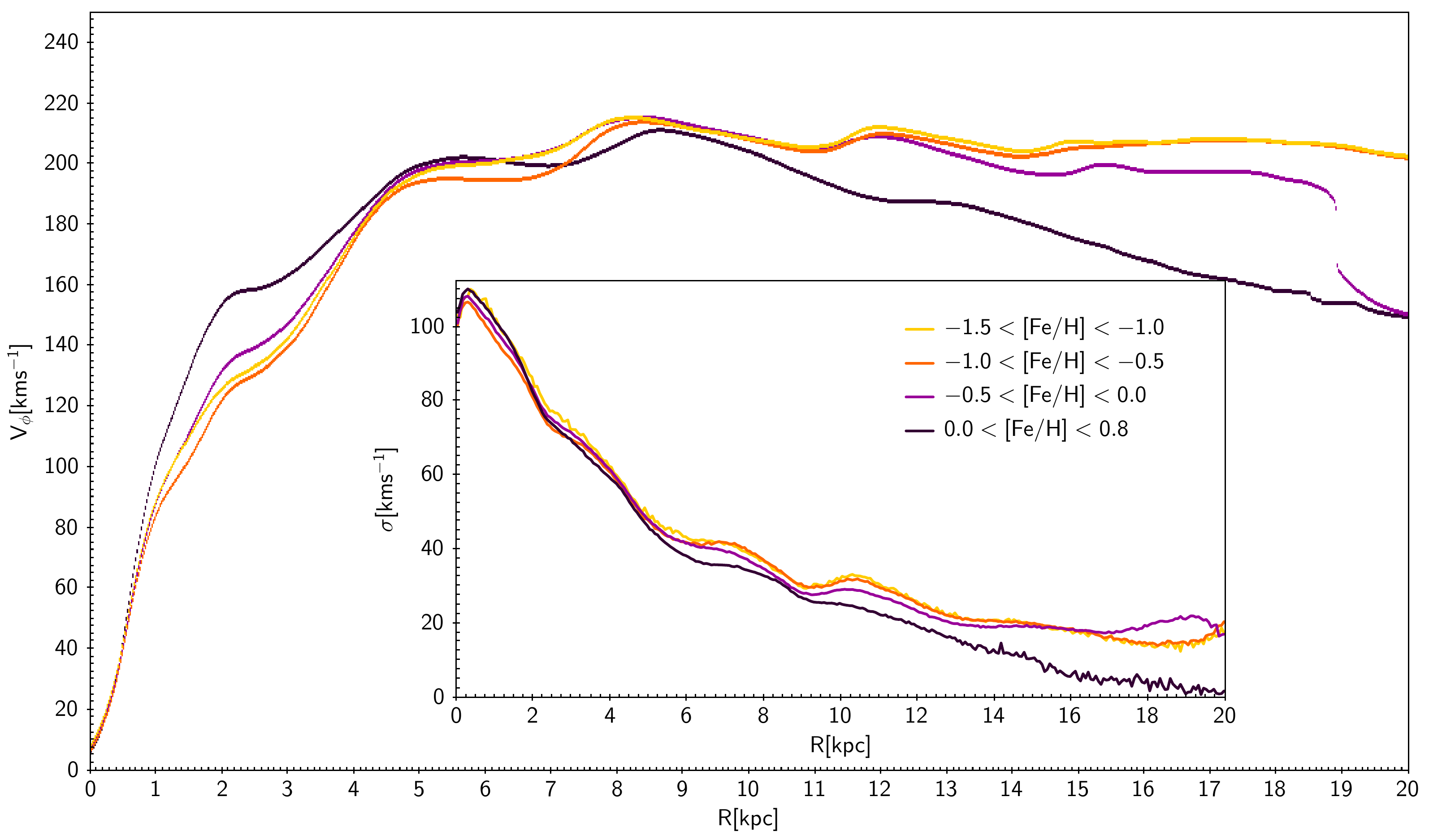}
\end{center}
\caption{Rotation velocity as a function of radius for populations with different metallicities.}
\label{Metallicity2}
\end{figure}

Figure \ref{Metallicity2} shows the RC of the disc splitted in stellar populations with different metallicitiess. The rotation within the inner 5kpc is higher for metal-rich stars and decreases for the metal-poor ones. Meanwhile, in the outer disc the trend is inverted. This is due to the specifics on the way the star formation rate, the age of the galactic chemical gradient, and the stellar migration interact. Also notice that at small radii the velocity dispersion is almost independent from the metallicity, but at the outer disc it decreases and is smaller for the metal-rich stars.

\subsection{Ridges, eccentricity, and wiggles}

In \citet{2019MNRAS.485L.104M} we found an explanation for the substructures in the form of ridges in the $V_{\Phi} - R$ plane revealed by the Gaia second Data Release (Gaia DR2) \citep{2018arXiv180410196A,2018MNRAS.479L.108K}. We showed that the diagonal ridges in velocity distribution are triggered by resonances of bar and spiral arms. Moreover, due to their layout, these ridges are projected as wiggles in the rotation curve of the disc.

\begin{figure}
\begin{center}
\includegraphics[width=\columnwidth]{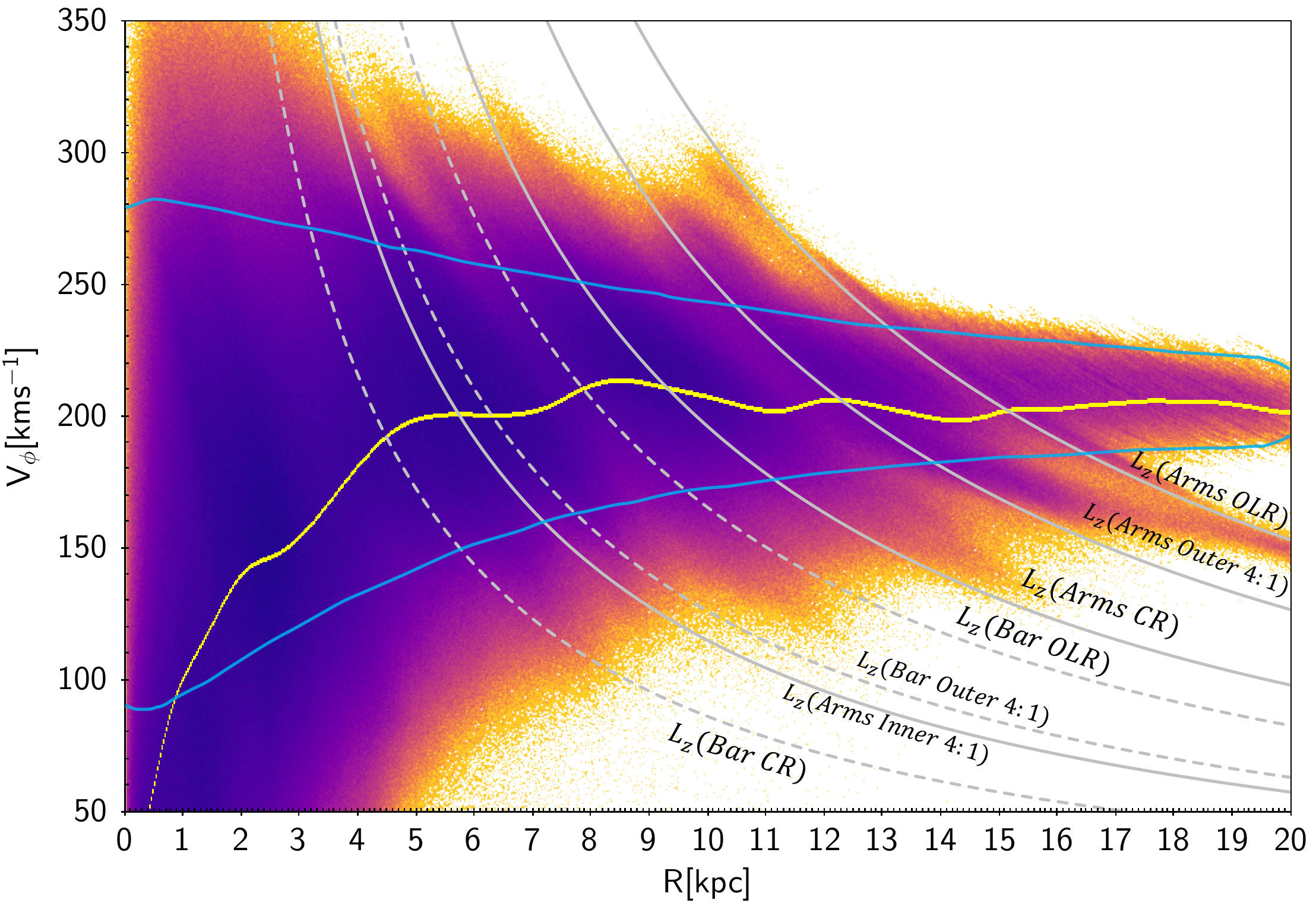}
\end{center}
\caption{Rotation velocity for stars in our MW simulation with bar+spirals. Ridges are found across the entire disc. Notice that these diagonal ridges project themselves in the form of wiggles on the RC (yellow line).The blue curve is the envelope of the distribution of stars for the axisymmetric case. Grey dashed and solid lines are $L_z$ curves associated with bar and spirals resonances, respectively.}
\label{ridgesB}
\end{figure}

Figure \ref{ridgesB} summarises our findings, it shows the $V_{\Phi}-R$ plane for the particles in our MW simulation: the blue line indicates the envelope of the distribution of stars in an equivalent model without bar and spirals. Notice that due to the presence of bar and spirals, the stellar velocities spread outside the blue envelope, mostly through diagonal ridges of specific angular momentum value $L_z$; moreover, such $L_z$ values are not arbitrary, they correspond to the angular momentum of the bar and spiral resonances. Figure \ref{ridgesB} shows that, for each family of orbits with constant $L_z$ value, this diffusion process increases the orbital apogalacticon and decreases the perigalacticon which in turns makes the orbits elongated and more eccentric. Hence, the diagonal extension of ridges in the $V_{\Phi}-R$ plane translates into substructures of high eccentricity in our eccentricity maps (Figs. \ref{Sa}--\ref{Sc} and Fig. \ref{MW}). Finally, the origin of eccentric zones in the disc is the same process that triggers the development of ridges in the $V_{\Phi}-R$ plane.

In this way the RC wiggles can be explained as imprints of orbital eccentricities, or as the projection of diagonal ridges in the $V_{\Phi}-R$ plane, and both explanations should be considered as two faces of the same mechanism.

\section{Discussion}

\subsection{The need of modelling wiggles and ridges}

Bumps and wiggles are ubiquitous features of RCs, revealing a perturbed state of the kinematics of the disc. It is clear that, to study RCs, it is necessary to go beyond the smooth components of the disc \citep[see for example the discussion by][]{mcgaugh19}. 

Several examples of galactic RCs, including that of our own galaxy, show wiggles with significantly high amplitude, which is consequence of the presence of non-circular motions of stars and gas. Moreover, \citet{mcgaugh19} finds kinematically inferred spiral arms from observations of the terminal velocities in the disc of the MW, which in turn correspond with the local variations in the density. This means that azimuthal and radial variations of the mass density are directly linked to the observed kinematics. With the recent use of satellites, the ridges of the $V_\phi$-$R$ plane have become apparent \citep{2018arXiv180410196A}. For these reasons, there is a clear need for more realistic (and elaborate) models of galactic structure in order to keep up with observations.

There are several ways to achieve wiggles and ridges in RCs and $V_\phi$-$R$ plane, respectively. \citet{mcgaugh19} himself presents an azimuthally averaged model to reproduce the wiggles in the MW's RC varying the density as a function of $R$. Beyond spiral arms + bars, there are a few additional mechanisms able to produce wiggles (and probably ridges): perturbations caused by dwarf galaxies crossing the galactic plane \citep{2015MNRAS.454..933D,2019MNRAS.485.3134L} or inherent oscillations in the mass density of the dark matter halo \citep{2014MNRAS.444..185M,2018MNRAS.475.1447B} are two such examples.

Yet, spiral arms are ubiquitous in spiral galaxies and their presence is enough to produce both wiggles and ridges. Bars alone are also able to produce ridges in the $V_\phi$-$R$ plane as well as wiggles in the inner parts of the RC \citep{2019MNRAS.485L.104M}. The combination of bar and spirals will produce a richer substructure in the velocity distribution of the stellar disc and the RCs.

\subsection{The nature of spiral arms}

We have shown that our own implementation of long-lasting spiral arms leaves an imprint on the galactic RCs in the form of bumps and wiggles, although the amplitude of this effect may vary for different natures of the arms.

The nature of spiral arms is not unique, leaving aside the grand bisymmetric spirals, stellar discs can also be perturbed by multiarm transient structures \citep{2013ApJ...766...34D}, flocculent structure with corotating features \citep{2012MNRAS.421.1529G}, spiral structure caused by perturbations from encounters with dwarf galaxies \citep{2015MNRAS.454..933D,2019MNRAS.485.3134L}, or the superposition of spiral patterns with a bar \citep{2010ApJ...722..112M, 2011MNRAS.417..762Q}.

Such diversity of non-axisymmetric structures disturb the stellar kinematics in the same general way: rearranging the angular momentum in the disc, displacing stars to outer and inner radii, and altering the orbital eccentricities. It is the amplitude of this effect will differ with the nature of the spiral arms, e.g., the RC could be greatly perturbed by strong arms triggered by the interaction of the disc with a dwarf galaxy, or  be barely disturbed by tightly wound, short-living, transient spirals.  

We will consider in a future study how the RC may look like under the influence of non-axisymmetric structures of different nature, for which it’s necessary to compare different models and simulations of spiral arms.

\section{Conclusions}
\label{conclusions}

We performed a comprehensive orbital study in observationally motivated three-dimensional galactic models. In order to analyze the dynamical effects of spiral arms and bar in the stellar disc we integrate several millions orbits, representing the stellar discs of an Sa, an Sb and an Sc galaxy, as well as the stellar disc of the specific case of the MW.

By separating the orbits according the their eccentricity value, we found that radially cold orbits trace better the differences in the rotation curve when it is measured along different angles in a galaxy that harbours non-axisymmetric structures, like bar and spiral arms.

Stars rotate slower along the spiral arms than along the inter-arm. This means that the rotation of the disc will have local minima or maxima when it is measured on an arm or inter-arm, respectively. Which in turn produces wiggles and bumps in the rotation curve.

Tightly wound arms ($\lesssim 10\deg$) do not have major influence in the orbital rearrangement of orbital eccentricities in the disc, making no significant dependence of the rotation as a function of the azimuth. On the other hand, wide open arms are able to induce major non-circular motions in the orbits, making their rotation speed to depend on the azimuthal position.

With the construction of the eccentricity maps for our galactic models we found that the kinematics of the disc is modified at the instantaneous position of the arms and can remain significantly altered in the regions swept by them; this produces substructures in the distribution of orbital eccentricities that may be reflected on its RC. 

RCs are also significantly altered by the presence of a central bar. The rotation speed measured parallel the bar is lower than the one measured perpendicular to it and they remain so for all radii inside the bar. This produces a lower or higher RC depending in which angle, with respect the bar, it is measured.

In spiral galaxies, not all stars are part of the disc and should not be considered to determine the disc RC, even stars that lie close to the mid-plane, may not trace properly the mean rotation of the disc. A measurement of the rotation curve near to the plane always carries a contribution from vertically hot orbits, specially for galaxies with relatively thick discs. 
When measuring RCs at different distances off the plane, we can see that the rotation is larger for stars with small observed vertical position, $|z_{obs}|$, and diminishes with increasing vertical distance. This effect is more important in early type galaxies.

The behaviour of the RCs as a function of maximum vertical amplitude $z_{max}$ is similar to the one observed in $|z_{obs}|$ (the rotation is faster for dynamically colder orbits, and this effect is more important for early type galaxies). The greatest difference is that the RCs as a function of $z_{max}$ are always above than the RCs as a function of $|z_{obs}|$; this is because at all radii the rotation of the disc decreases as the altitude increases.

The Gaia DR2 revealed a richness of substructures in the kinematics of the MW disc. For instance, we are aware of the distinctive pattern of diagonal ridges covering the $V_{\phi}-R$ plane \citep{2018arXiv180410196A,2018MNRAS.479L.108K}. In \citep{2019MNRAS.485L.104M} we found that these ridges have a resonant origin, induced by arms and bar, and manifest themselves as bumps and wiggles in the rotation curve (see Figure \ref{ridgesB}). Here we further notice that the development of diagonal ridges in the $V_{\phi}-R$ plane and the development of high orbital eccentricities in the stellar disc are the same. The diagonal extension of the ridges causes the families of orbits to increase their apogalacticon and decrease their perigalacticon, which in turn increases their orbital eccentricity. Hence the following explanations of bumps and wiggles in RCs are equivalent: they are manifestations of diagonal ridges in the $V_{\phi}-R$ plane, or of the rearrangement of the orbital eccentricities in the stellar disc.

\section*{Acknowledgements}
This paper is dedicated to the loving memory of B\'arbara Pichardo, who was deeply involved in the research and writing of this work, her active mind and presence are missed. We would like to acknowledge an anonymous referee for a careful review and some insightful suggestions. We acknowledge DGTIC-UNAM for providing HPC resources on the Cluster Supercomputer Miztli. The authors acknowledge support from CONACyT Ciencia B\'asica grant 255167. L.A.M.M. and B.P. acknowledge support from PAPIIT IN-101918. A.P. acknowledges support from PAPIIT IG-100319 grant. 
 

\label{lastpage}


\begin{thebibliography}{101}

\bibitem[Allen \& Santill\'an(1991)]{AS91}
Allen C., Santill\'an A., 1991, \rmxaa, 22, 255

\bibitem[Antoja et al.(2009)]{2009ApJ...700L..78A} Antoja, T., Valenzuela, O., Pichardo, B., et al.\ 2009, \apjl, 700, L78 

\bibitem[Antoja et al.(2018)]{2018arXiv180410196A} Antoja, T., Helmi, A., Romero-G{\'o}mez, M., et al.\ 2018, \nat, 561, 360 

\bibitem[Begeman(1987)]{1987PhDT.......199B} Begeman, K.~G.\ 1987, Ph.D.~Thesis, 

\bibitem[Bernal et al.(2018)]{2018MNRAS.475.1447B} Bernal, T., Fern{\'a}ndez-Hern{\'a}ndez, L.~M., Matos, T., et al.\ 2018, \mnras, 475, 1447

\bibitem[Begeman(1989)]{1989A&A...223...47B} Begeman, K.~G.\ 1989, \aap, 223, 47 

\bibitem[Blitz(1979)]{Blitz79} Blitz, L.\ 1979, \apjl, 231, L115

\bibitem[Block \& Puerari(1999)]{1999A&A...342..627B} Block, D.~L., \& Puerari, I.\ 1999, \aap, 342, 627 

\bibitem[Bobylev et al.(2008)]{Bovylev08} Bobylev, V.~V., Bajkova, A.~T., \& Stepanishchev, A.~S.\ 2008, Astronomy Letters, 34, 515

\bibitem[Bond et al.(2010)]{2010ApJ...716....1B} Bond, N.~A., Ivezi{\'c}, {\v Z}., Sesar, B., et al.\ 2010, \apj, 716, 1 

\bibitem[Bosma(1981a)]{1981AJ.....86.1791B} Bosma, A.\ 1981a, \aj, 86, 1791 

\bibitem[Bosma(1981b)]{1981AJ.....86.1825B} Bosma, A.\ 1981b, \aj, 86, 1825 

\bibitem[Bovy et al.(2012)]{Bovy_etal2012} Bovy, J., Allende Prieto, C., Beers, T.~C., et al.\ 2012, \apj, 759, 131

\bibitem[Brand \& Blitz(1993)]{Brand93} Brand, J., \& Blitz, L.\ 1993, \aap, 275, 67

\bibitem[Chemin et al.(2015)]{2015A&A...578A..14C} Chemin, L., Renaud, F., \& Soubiran, C.\ 2015, \aap, 578, A14 

\bibitem[Clemens(1985)]{Clemens85} Clemens, D.~P.\ 1985, \apj, 295, 422

\bibitem[Dicaire et al.(2008)]{Dicaire08} Dicaire, I., Carignan, C., Amram, P., et al.\ 2008, \mnras, 385, 

\bibitem[de la Vega et al.(2015)]{2015MNRAS.454..933D} de la Vega, A., Quillen, A.~C., Carlin, J.~L., et al.\ 2015, \mnras, 454, 933

\bibitem[D'Onghia et al.(2013)]{2013ApJ...766...34D} D'Onghia, E., Vogelsberger, M., \& Hernquist, L.\ 2013, \apj, 766, 34

\bibitem[Faure et al.(2014)]{2014MNRAS.440.2564F} Faure, C., Siebert, A., \& Famaey, B.\ 2014, \mnras, 440, 2564 

\bibitem[Fich et al.(1989)]{Fich89} Fich, M., Blitz, L., \& Stark, A.~A.\ 1989, \apj, 342, 272

\bibitem[Freudenreich(1998)]{1998ApJ...492..495F} Freudenreich,
  H.~T.\ 1998, \apj, 492, 495

\bibitem[Gilmore et al.(2002)]{2002ApJ...574L..39G} Gilmore, G., Wyse, R.~F.~G., \& Norris, J.~E.\ 2002, \apjl, 574, L39 

\bibitem[Grand et al.(2012)]{2012MNRAS.421.1529G} Grand, R.~J.~J., Kawata, D., \& Cropper, M.\ 2012, \mnras, 421, 1529

\bibitem[Guhathakurta et al.(1988)]{1988AJ.....96..851G} Guhathakurta, P., van Gorkom, J.~H., Kotanyi, C.~G., \& Balkowski, C.\ 1988, \aj, 96, 851 

\bibitem[Honma \& Sofue(1997)]{Honma97} Honma, M., \& Sofue, Y.\ 1997, \pasj, 49, 453

\bibitem[Kawata et al.(2018)]{2018MNRAS.479L.108K} Kawata, D., Baba, J., Ciuc{\v a}, I., et al.\ 2018, \mnras, 479, L108 

\bibitem[Laine et al.(1999)]{1999MNRAS.302L..33L} Laine, S., Knapen, J.~H., Perez-Ramirez, D., Doyon, R., \& Nadeau, D.\ 1999, \mnras, 302, L33 

\bibitem[Laporte et al.(2019)]{2019MNRAS.485.3134L} Laporte, C.~F.~P., Minchev, I., Johnston, K.~V., et al.\ 2019, \mnras, 485, 3134

\bibitem[L{\'e}pine et al.(2011)]{2011MNRAS.417..698L} L{\'e}pine, J.~R.~D., Cruz, P., Scarano, S., Jr., et al.\ 2011, \mnras, 417, 698 

\bibitem[Leung et al.(2018)]{2018MNRAS.477..254L} Leung, G.~Y.~C., Leaman, R., van de Ven, G., et al.\ 2018, \mnras, 477, 254 

\bibitem[L{\'o}pez-Corredoira(2014)]{Lopez_Corredoira14} L{\'o}pez-Corredoira, M.\ 2014, \aap, 563, A128 

\bibitem[Martinez-Medina \& Matos(2014)]{2014MNRAS.444..185M} Martinez-Medina, L.~A., \& Matos, T.\ 2014, \mnras, 444, 185

\bibitem[Martinez-Medina et al.(2015)]{2015ApJ...802..109M}
  Martinez-Medina, L.~A., Pichardo, B., P{\'e}rez-Villegas, A., \&
  Moreno, E.\ 2015, \apj, 802, 109

\bibitem[Martinez-Medina et al.(2017)]{2017MNRAS.468.3615M} Martinez-Medina, L.~A., Pichardo, B., Peimbert, A., \& Carigi, L.\ 2017, \mnras, 468, 3615 

\bibitem[Martinez-Medina et al.(2018)]{2018MNRAS.474...32M} Martinez-Medina, L.~A., Gieles, M., Pichardo, B., \& Peimbert, A.\ 2018, \mnras, 474, 32

\bibitem[Martinez-Medina et al.(2019)]{2019MNRAS.485L.104M} Martinez-Medina, L., Pichardo, B., Peimbert, A., et al.\ 2019, \mnras, 485, L104

\bibitem[McGaugh(2019)]{mcgaugh19} McGaugh, S.~S.\ 2019, \apj, 885, 87

\bibitem[Merrifield(1992)]{Merrifield1992} Merrifield, M.~R.\ 1992, \aj, 103, 1552

\bibitem[Minchev \& Famaey(2010)]{2010ApJ...722..112M} Minchev, I., \& Famaey, B.\ 2010, \apj, 722, 112

\bibitem[Minchev et al.(2012)]{2012A&A...548A.126M} Minchev, I., Famaey, B., Quillen, A.~C., et al.\ 2012, \aap, 548, A126 

\bibitem[Miyamoto \& Nagai(1975)]{1975PASJ...27..533M} Miyamoto, M., \& Nagai, R.\ 1975, \pasj, 27, 533 

\bibitem[Moreno et al.(2015)]{2015MNRAS.451..705M} Moreno, E.,
  Pichardo, B., \& Schuster, W.~J.\ 2015, \mnras, 451, 705 

\bibitem[Pease(1918)]{1918PNAS....4...21P} Pease, F.~G.\ 1918, Proceedings of the National Academy of Science, 4, 21

\bibitem[P{\'e}rez-Villegas et al.(2012)]{2012ApJ...745L..14P} P{\'e}rez-Villegas, A., Pichardo, B., Moreno, E., Peimbert, A., \& Vel{\'a}zquez, H.~M.\ 2012, \apjl, 745, L14 

\bibitem[P{\'e}rez-Villegas et al.(2013)]{2013ApJ...772...91P} P{\'e}rez-Villegas, A., Pichardo, B., \& Moreno, E.\ 2013, \apj, 772, 91

\bibitem[P{\'e}rez-Villegas et al.(2015a)]{2015MNRAS.451.2922Pa} P{\'e}rez-Villegas, A., G{\'o}mez, G.~C., \& Pichardo, B.\ 2015a, \mnras, 451, 2922

\bibitem[P{\'e}rez-Villegas et al.(2015b)]{2015ApJ...809..170Pb} P{\'e}rez-Villegas, A., Pichardo, B., \& Moreno, E.\ 2015b, \apj, 809, 170

\bibitem[Persic et al.(1996)]{1996MNRAS.281...27P} Persic, M., Salucci, P., \& Stel, F.\ 1996, \mnras, 281, 27 

\bibitem[\protect\citeauthoryear{Pichardo et al.}{2003}]{PMME03}
  Pichardo, B., Martos, M., Moreno, E. \& Espresate, J., 2003, ApJ,
  582, 230

\bibitem[Pichardo et al.(2004)]{Pichardo2004} Pichardo, B., Martos, 
M., \& Moreno, E.\ 2004, \apj, 609, 144

\bibitem[Pichardo et al.(2012)]{2012AJ....143...73P} Pichardo, B.,
  Moreno, E., Allen, C., et al.\ 2012, \aj, 143, 73

\bibitem[Quillen et al.(2011)]{2011MNRAS.417..762Q} Quillen, A.~C., Dougherty, J., Bagley, M.~B., Minchev, I., \& Comparetta, J.\ 2011, \mnras, 417, 762 

\bibitem[Randriamampandry et al.(2016)]{2016A&A...594A..86R} Randriamampandry, T.~H., Deg, N., Carignan, C., Combes, F., \& Spekkens, K.\ 2016, \aap, 594, A86 

\bibitem[Reid et al.(2014)]{Reid14} Reid, M.~J., Menten, K.~M., Brunthaler, A., et al.\ 2014, \apj, 783, 130

\bibitem[Roberts \& Rots(1973)]{1973A&A....26..483R} Roberts, M.~S., \& Rots, A.~H.\ 1973, \aap, 26, 483 

\bibitem[Roelfsema \& Allen(1985)]{1985A&A...146..213R} Roelfsema, P.~R., \& Allen, R.~J.\ 1985, \aap, 146, 213 

\bibitem[Ro{\v s}kar et al.(2012)]{2012MNRAS.426.2089R} Ro{\v s}kar, R., 
Debattista, V.~P., Quinn, T.~R., \& Wadsley, J.\ 2012, \mnras, 426, 2089 

\bibitem[Rubin \& Ford(1970)]{1970ApJ...159..379R} Rubin, V.~C., \& Ford, W.~K., Jr.\ 1970, \apj, 159, 379 

\bibitem[Rubin et al.(1978)]{1978ApJ...225L.107R} Rubin, V.~C., Ford, W.~K., Jr., \& Thonnard, N.\ 1978, \apjl, 225, L107 

\bibitem[Rubin et al.(1985)]{1985ApJ...289...81R} Rubin, V.~C., Burstein, D., Ford, W.~K., Jr., \& Thonnard, N.\ 1985, \apj, 289, 81 

\bibitem[Sellwood(2014)]{2014RvMP...86....1S} Sellwood, J.~A.\ 2014, Reviews of Modern Physics, 86, 1 
  
\bibitem[Shetty et al.(2007)]{2007ApJ...665.1138S} Shetty, R., Vogel, S.~N., Ostriker, E.~C., \& Teuben, P.~J.\ 2007, \apj, 665, 1138 

\bibitem[Slipher(1914)]{1914LowOB...2...66S} Slipher, V.~M.\ 1914, Lowell Observatory Bulletin, 2, 66 

\bibitem[Sofue et al.(1999)]{1999ApJ...523..136S} Sofue, Y., Tutui, Y., Honma, M., et al.\ 1999, \apj, 523, 136 

\bibitem[Sofue \& Rubin(2001)]{2001ARA&A..39..137S} Sofue, Y., \& Rubin, V.\ 2001, \araa, 39, 137 

\bibitem[Sofue et al.(2009)]{Sofue09} Sofue, Y., Honma, M., \& Omodaka, T.\ 2009, \pasj, 61, 227

\bibitem[Spekkens \& Sellwood(2007)]{Spekkens_Sellwood07} Spekkens, K., \& Sellwood, J.~A.\ 2007, \apj, 664, 204

\bibitem[Valenzuela et al.(2007)]{Valenzuela07} Valenzuela, O., Rhee, G., Klypin, A., et al.\ 2007, \apj, 657, 773

\bibitem[van Albada et al.(1985)]{1985ApJ...295..305V} van Albada, T.~S., Bahcall, J.~N., Begeman, K., \& Sancisi, R.\ 1985, \apj, 295, 305 

\bibitem[Williams et al.(2013)]{2013MNRAS.436..101W} Williams, M.~E.~K., Steinmetz, M., Binney, J., et al.\ 2013, \mnras, 436, 101 

\bibitem[Wong et al.(2004)]{2004ApJ...605..183W} Wong, T., Blitz, L., \& Bosma, A.\ 2004, \apj, 605, 183 

\end{thebibliography}
\end{document}